\documentclass{article}
\usepackage{spconf,amsmath,graphicx}
\usepackage{booktabs}
\usepackage{multirow}
\usepackage[vlined,linesnumbered,ruled]{algorithm2e}
\usepackage{xcolor}
\usepackage{amsfonts}
\usepackage{etoolbox}

\newcommand{\Anurag}[1]{{\color{red}{\bf Anurag: }#1}} 

\def\X{{\mathbf X}}
\def\w{{\mathbf w}}
\def\D{{\mathbf D}}
\def\P{{\mathbf P}}

\newcommand{\ra}[1]{\renewcommand{\arraystretch}{#1}}

\usepackage[
backend=biber,
style=ieee,
doi=false,isbn=false,url=false,eprint=false
]{biblatex}

\defbibheading{bibliography}[\refname]{}

\apptocmd{\thebibliography}{\setlength{\itemsep}{0.1pt}}{}{}

\title{PAAPLoss: A Phonetic-Aligned Acoustic Parameter Loss for Speech Enhancement}
\name{\begin{tabular}{c}Muqiao Yang$^1$, Joseph Konan$^1$, David Bick$^1$, Yunyang Zeng$^1$, Shuo Han$^1$, \\Anurag Kumar$^2$, Shinji Watanabe$^1$, Bhiksha Raj$^1$ \end{tabular}}

\address{   
$^1$ Carnegie Mellon University, $^2$Meta Reality Labs Research}
\begin{document}
\ninept
\maketitle
\begin{abstract}


Despite rapid advancement in recent years, current speech enhancement models often produce speech that differs in perceptual quality from real clean speech. We propose a learning objective that formalizes differences in perceptual quality, by using domain knowledge of acoustic-phonetics. We identify temporal acoustic parameters -- such as spectral tilt, spectral flux, shimmer, etc. -- that are non-differentiable, and we develop a neural network estimator that can accurately predict their time-series values across an utterance. We also model phoneme-specific weights for each feature, as the acoustic parameters are known to show different behavior in different phonemes. We can add this criterion as an auxiliary loss to any model that produces speech, to optimize speech outputs to match the values of clean speech in these features. Experimentally we show that it improves speech enhancement workflows in both time-domain and time-frequency domain, as measured by standard evaluation metrics. We also provide an analysis of phoneme-dependent improvement on acoustic parameters, demonstrating the additional interpretability that our method provides. This analysis can suggest which features are currently the bottleneck for improvement.



\end{abstract}
%


\section{Introduction}
\label{sec:intro}
Speech enhancement (SE) tries to extract clean speech from signals that have been degraded mainly by noise. The ability to remove noise from speech is extremely useful, as noisy environments commonly affect applications such as VoIP and phone calls, hearing aids, and downstream speech processing tasks. Our focus is on the more ubiquitous single-channel speech enhancement which does not require multi-microphone speech capture.

In the last decade, single-channel SE has greatly improved by moving from traditional signal processing techniques to deep neural networks (DNN) \cite{cnn_denoising, rnn_denoising, regression_dnn_denoising, Weninger2015SpeechEW, plantinga2020phonetic}.  Deep Noise Suppression (DNS) challenges have further stimulated single-channel SE work by providing a large corpus of audio synthesized over a wide range of noise types and levels \cite{reddy2020interspeech, dubey2022icassp}. It also provides a common test set to measure performance.

 Single-channel SE models are usually trained by comparing enhanced speech to clean speech using point-wise differences between waveforms or spectrograms. While this paradigm has been effective, SE models often still generate unnatural sounding speech \cite{dnsmos_pesq_flaw}. Limitations with these classic losses include failure to capture pitch \cite{turian_henry}, and relatively low improvement for low-energy phonemes \cite{plantinga2021perceptual}. Additionally, \cite{perceptual_jnd} and \cite{metric_gan_plus} describe that $\ell_1$ or $\ell_2$ difference at the signal level is not highly correlated with speech quality.  
 
 Other approaches have sought to address these issues, including optimization of perceptual evaluation metrics. However, these are non-differentiable, so approximations offer limited improvements \cite{white_box_perceptual_loss, metric_RL_SE}, require cumbersome optimization \cite{metricgan, metric_RL_SE} or offer little to no interpretability through domain knowledge \cite{hsieh2020improving}. We aim to address these problems in this paper and we try to accomplish it by incorporating domain knowledge through fundamental speech features which we refer to as \emph{acoustic parameters}.
 
 
 
  Before the rise of DNNs, features such as pitch, jitter, shimmer, spectral tilt -- to name a few -- were used as inputs to shallow models, such as in speaker and emotion recognition \cite{egemaps}. They lost popularity as DNNs gained more success operating directly on waveforms or spectrograms. Their non-differentiable computations also inhibit their straightforward use in optimization of DNNs. Nevertheless, these parameters provide critical information about frequency content, energy/amplitude, and other spectral qualities of the speech signal. Prior perceptual studies have shown important associations of these features to voice quality \cite{correlates_breathy_rough, correlates_breathy, KASUYA1986171}. \cite{yang2022improving} introduced a differentiable estimator of utterance-level statistics for these parameters and improved state-of-the-art SE models through an auxiliary loss aimed to minimize the differences between parameter values of clean and enhanced speech. Similarly, we work with 25 acoustic parameters enumerated in the extended Geneva Minimal Acoustic Parameter Set \cite{egemaps}. However, unlike prior work which considered these acoustic parameters at the global (utterance) level through summary statistics, we incorporate the temporal aspects of these acoustic parameters \cite{taploss}. Furthermore, we incorporate the associations between acoustic parameters and phonemes which have been studied previously in the sub-field of acoustic-phonetic. For example, plosives typically have a high amplitude followed by a very low amplitude, as they are produced by complete closure in the vocal tract followed by a sudden release of pressure \cite{alwan2011perception}. Nasality in sounds introduces anti-formants because the nasal cavity introduces resonances that interfere with the resonances of the vocal tract \cite{nasals}. Each vowel also has different formant structures based on the resonances created by different locations of constriction in the vocal tract \cite{yi2019encoding}. 
 
 In this paper, we introduce a phonetic-aligned acoustic parameter (PAAP) loss to improve speech outputs from SE systems. We accomplish this by minimizing the difference between phonetically-aligned acoustic parameters in enhanced speech and clean speech. This is done with a two-step approach. First, we introduce a differentiable estimator of temporal acoustic parameters, to obtain the time series of each parameter across an utterance. Second, we calculate differentiable phoneme-specific weights for each acoustic parameter based on their ability to predict phoneme logits. This allows us to put different emphases on acoustic parameters at one time step, depending on the predicted phoneme at the same time step. These two components allow us to optimize the original model end-to-end to match clean speech with phonetic-aligned acoustic parameters. Our approach leads to improvements over competitive SE models. More importantly though, we demonstrate the interpretability of our method, by analyzing the phoneme-dependent improvement on acoustic parameters. 


\vspace{-2mm}
\section{Related work}
\label{sec:related}


Various works have tried to introduce losses aimed at improving the perceptual quality. Some techniques include optimization of non-differentiable perceptual metrics through generative adversarial networks (GAN) \cite{metricgan}, reinforcement learning \cite{metric_RL_SE}, and convex approximations of metrics \cite{white_box_perceptual_loss}. However, as shown in \cite{yang2022improving}, current methods fail to capture the aforementioned acoustic parameters, and explicit supervision of retaining them improved model outputs. 

Other methods have attempted to use phonetic information in enhancing perceptual quality, such as \cite{hsieh2020improving}. However, their loss function did not explicitly use domain knowledge of phonemes and the phonetic information was only implicitly captured in wav2vec embeddings. Recently, \cite{tal22_interspeech} performed a study of phonetic-aware techniques for speech enhancement but relies on uninterpretable HuBERT features \cite{hsu2021hubert}. Both techniques are evaluated on the Valentini dataset, which is much smaller and less varied than in our experiments. Moreover, our method allows interpretability through both acoustic parameters and phonemes, as illustrated in the experiments section. Lastly, \cite{yang2022improving} also used the acoustic parameters for optimization of perceptual quality. However, it did not factor in \textit{temporal} or \textit{phonetic} information. As these acoustic parameters vary greatly over an utterance, and between phonemes, modeling this phoneme and temporal dependencies can be helpful for improved performance.

\vspace{-2mm}
\section{Method}
\label{sec:method}

 
We propose to use a phonetic-aligned acoustic parameter loss to fine-tune SE models. Note that this objective function can be applied to any architecture, and even any task that involves speech outputs. In this section we describe the use in SE as a concrete example. However it only requires a waveform as input, and it is end-to-end differentiable, so the PAAP Loss can be applied to any model that produces waveform.  

The overall learning paradigm is summarized in Algorithm \ref{algo:algo}. We will present the temporal acoustic parameter estimation in Subsection \ref{subsec:temporal-ap-estimation}, the phonetic-alignment and weighting in Subsection \ref{subsec:phonetic-alignment}, and the overall fine-tuning process with the proposed PAAP Loss in Subsection \ref{subsec:paap-loss}.  

\vspace{-2mm}
\subsection{Temporal Acoustic Parameter Estimation}
\label{subsec:temporal-ap-estimation}


 First, we take the pre-trained SE model as our seed model $\Phi$, and pass in the noisy audio $\X^N$ to obtain the enhanced waveform $\X^E$ (line 3). On top of the seed models, we use a pre-trained estimator network $\Psi$ to predict the acoustic parameters given a raw waveform. The acoustic parameters include a set of 25 low-level descriptors, covering prosodic, excitation, vocal tract, and spectral descriptors that are found to be the most expressive of the acoustic characteristics as standardized feature set. 
 
 Unlike prior work which models these acoustic parameters at the utterance level through summary statistics, we incorporate the temporal feature of these acoustic parameters in the modeling. We pass the enhanced and clean waveforms to the model to predict temporal acoustic parameter matrices, $\D^E$ and $\D^C$ respectively (lines 4-5). The estimator network first performs short-time Fourier Transform (STFT) on the raw waveform, and then passes the spectrogram to a sequential neural network to obtain the predicted temporal acoustic parameters. We note that using the estimated clean acoustic parameters in PAAP Loss rather than ground-truth allows much greater ease of use by other researchers, as they do not have to synthesize labels from another toolkit, as in \cite{yang2022improving}. This benefits us by making our loss more accessible in an arbitrary SE network.
 
 

\SetKwComment{Comment}{// }

\newcommand\mycommfont[1]{\footnotesize\ttfamily\textcolor{black}{#1}}
\SetCommentSty{mycommfont}

\begin{algorithm}[!t]
\KwSty{Input: } Noisy waveform $\X^N$, clean waveform $\X^C$, seed model $\Phi$, pre-trained acoustic low-level descriptor estimator $\Psi$, estimated acoustic-phonetic weights $\w$.

\KwSty{Output: } calculated PAAP Loss $\ell_{\text{PAAP}}$

$\X^E \gets \Phi(\X^N)$  \tcp*[r]{\it Enhanced waveform from current model}

$\D^C \gets \Psi(\X^C)$  \tcp*[r]{\it Estimated clean acoustic parameters}

$\D^E \gets \Psi(\X^E)$  \tcp*[r]{\it Estimated enhanced acoustic parameters}

$\ell_{\text{PAAP}} \gets 0$ 

$N \gets \mathrm{len}(\X^C)$ \tcp*[r]{\it the total number of frames} 

\For{$i \gets 1$ \KwTo $N$ }{
$j \gets$ Index of phoneme at $\X^{C}_{i}$

$\ell_{\text{PAAP}} \gets \ell_{\text{PAAP}} + ( \D^{E}_i - \D^{C}_i )^2 \cdot \w_j$
}


$\ell_{\text{PAAP}} \gets \frac{1}{N} \cdot  \ell_{\text{PAAP}}$ 

return $\ell_{\text{PAAP}}$

\caption{Overall workflow of applying PAAP Loss in one iteration of our SE paradigm.}
\label{algo:algo}
\end{algorithm}

\subsection{Phonetic Alignment}
\label{subsec:phonetic-alignment}
The next component of the PAAP Loss is the set of acoustic-phonetic weights $\w$, as we would like to weigh the acoustic parameters differently based on their importance to predict phoneme logits. These acoustic-phonetic weights are estimated using clean speech, through linear regression between the acoustic parameters and their corresponding segmented phoneme logits:
\begin{equation}
    \w = ((\D^C)^\top \D^C))^{-1}((\D^C)^\top \P^C) \\
    \vspace{-1mm}
\end{equation}

where $\P^C$ indicates the phoneme logits of the clean waveform. Each column $\w_i$ is the vector of weights from the 25 acoustic parameters to phoneme $i$, plus a bias term. Each weight $\w_{ij}$ corresponds to how much a unit change in acoustic parameter $i$ changes the log-probability of phoneme $j$. 

The weights reflect how much information each feature contains about each phoneme, so we can use it to emphasize optimization on differences between clean and enhanced parameter values that are more significant for the current phoneme.

We obtain $\P^C$ using an unsupervised phonetic aligner with a vocabulary of 40 phonemes, and one index for silence. We retain the silence index as we expect the relationship between acoustic parameters and phonemes will be different over non-speech regions of the utterance, and we would like to include this in the modeling. The unsupervised phonetic aligner allows flexibility to apply our method on datasets without ground-truth transcriptions. 

\subsection{Fine-tuning with PAAP Loss}
\label{subsec:paap-loss}

During fine-tuning, we first predict the phoneme index $j$ for each frame across time, using the argmax of predicted phoneme logits from clean audio. We will then use $\w_j$, the acoustic-phonetic weight for phoneme $j$. We calculate the squared difference between the clean and enhanced acoustic parameters at the current time step, and then perform dot-product with $\w_j$ (line 8-10). Note that these weights are used to incorporate phonetic information in the acoustic parameter differences, not to directly predict phoneme logits. In this way, the PAAP Loss calculates the weighted difference between acoustic parameters for each time step.  


In our implementation, we use STFT with hop length of $160$ and window length of $512$ to determine the total number of frames $N$. Both the phoneme logits and acoustic parameters have $N$ vectors of values. We iterate the above process over all frames in the utterance, and average the PAAP Loss by the total number of frames. The PAAP Loss is used as an auxiliary loss alongside the original loss of the SE model to fine-tune the network. We follow the optimal setting of \cite{yang2022improving} by keeping all weights frozen except the speech enhancement model. In our work, this applies to both acoustic-phonetic weights $\w$ and the weights of the temporal acoustic estimator network $\Psi$.



\vspace{-2mm}
\section{Experiments}
\label{sec:exp}

\begin{table}[t]
    \centering
    \hspace{-5mm}
    \ra{0.9}
    \begin{tabular}{c|c|cc|cc}
        \toprule
        \multirow{3}{*}{Metrics} & \multirow{3}{*}{Noisy} & \multicolumn{2}{c|}{FullSubNet} & \multicolumn{2}{c}{Demucs} \\
         & & \multirow{2}{*}{Baseline} & \multirow{2}{*}{\shortstack{PAAP\\Loss}} & \multirow{2}{*}{Baseline} & \multirow{2}{*}{\shortstack{PAAP\\Loss}} \\
        & & & & & \\
        \midrule
        PESQ ($\uparrow$) & 1.58 & 2.89 & \textbf{3.00} & 2.65 & \textbf{2.99} \\
        STOI ($\uparrow$) & 91.52 & 96.41 & \textbf{96.70} & 96.54 & \textbf{97.12} \\
        DNSMOS ($\uparrow$) & 2.48 & 3.21 & \textbf{3.27} & 3.31 & \textbf{3.34}  \\
        NORESQA ($\uparrow$) & 2.92 & 4.08  & \textbf{4.13} & 3.93 & \textbf{3.99}  \\
        WER ($\downarrow$) & 19.0 & 12.6 & \textbf{12.1} & 15.0 & \textbf{13.2} \\
        \bottomrule
    \end{tabular}
    \caption{Evaluation results of using the PAAP Loss compared with noisy audios and baseline models on the synthetic test set.}
    \label{tab:result}
    \vspace{-5mm}
\end{table}

\subsection{Data}
\label{subsec:data}
We used data and scripts from the Deep Noise Suppression (DNS) Challenge from InterSpeech 2020 \cite{reddy2020interspeech} to synthesize 50,000 pairs of 30-second (s) noisy and clean audio for training. We further synthesized another 10,000 audio pairs for validation set. The synthesis is performed under the default setting, where the Signal to Noise Ratio (SNR) is sampled uniformly between 0 and 40 decibels (dB). Then, noise audios from DNS noise set are selected with sufficient duration to span the selected clean utterance from Librivox, and added to the clean \cite{gemmeke2017audio}.

Our baseline models pre-process their input data slightly before training, and we follow each model's respective configuration during its fine-tuning. Demucs splits 30s audios into 10s segments with a 2s stride, and FullSubNet randomly samples a 3.072s segment from the 30s audio during each iteration. 

For the final evaluation of the models, we use the DNS 2020 synthetic test set with no reverberation. This set consists of 150 utterances from Graz University's clean speech dataset \cite{graz_university}, combined with noise categories randomly sampled from more than 100 noise classes. The SNR levels of the test set were uniformly sampled between 0 and 25 dB.

\subsection{Experimental Results}
To demonstrate that our proposed method is robust at improving various architectures, we select state-of-the-art Demucs \cite{defossez2020real} and FullSubNet \cite{hao2021fullsubnet} representing time domain and time-frequency domain models, respectively. These models are also open-sourced, so we use their pre-trained checkpoints to allow the reproducibility of the results of our work. For our unsupervised phonetic aligner, we use a wav2vec2-based method \cite{zhu2022phone}.

In our experiments, we weigh the PAAP Loss by a factor of $0.1$ before adding to the original loss to fine-tune the seed SE model. In the fine-tuning process, the pre-trained temporal acoustic parameter estimator is a 3-layer bi-directional long short-term memory (LSTM) \cite{LSTM} with $512$ hidden units. Table \ref{tab:result} shows the evaluation results evaluation by fine-tuning FullSubNet and Demucs with the additional PAAP Loss. 

We first look at Perceptual Evaluation of Speech
Quality (PESQ) and Short-Time Objective Intelligibility
(STOI) as they are canonical evaluations for speech enhancement. We see significant improvements in these metrics using our PAAP loss. Note that these are strong state-of-the-art models and hence improvements are hard to achieve. PESQ in particular improves by almost \textbf{4\%} and \textbf{13\%} for FullSubNet and Demucs respectively. 


Since our goal is to improve perceptual quality, the gold standard evaluation is mean opinion score from humans. This is calculated as the average of ratings on a 1-5 scale. Conducting a Mean Opinion Score (MOS) study is costly so we include two of the current state-of-the-art estimation approaches to estimate MOS, DNSMOS \cite{reddy2022dnsmos}, and NORESQA-MOS (Non-matching Reference based Speech Quality Assessment) \cite{manocha2022noresqa}. We observe that our PAAP loss once again shows improvements in these metrics for both models. 

Finally, we also calculate word error rate (WER) to evaluate whether our enhancement reduces distortions that affect downstream speech processing applications. Since we do not have ground-truth transcriptions, we use WavLM \cite{chen2022wavlm} base model from HuggingFace on clean speech to get the transcriptions as reference. We then apply the same recognizer to baseline and our enhanced speech to compare. We see improvements in WER as well, demonstrating that our method benefits both human perceptual quality and the ability to interface with speech technologies.  



\begin{figure}[t]
    \centering
    \hspace*{-0.1in}
    \includegraphics[width=1.13\columnwidth]{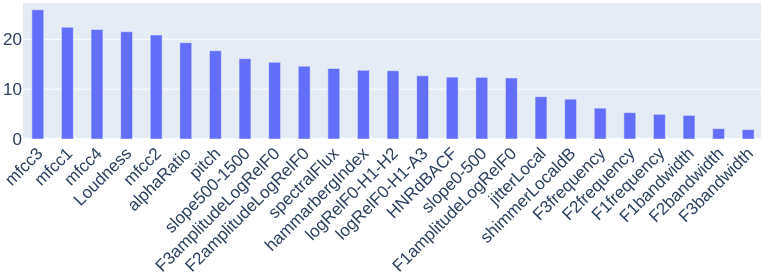} \\
    \vspace{1mm}
    \hspace*{-0.2in}
    \includegraphics[width=1.16\columnwidth]{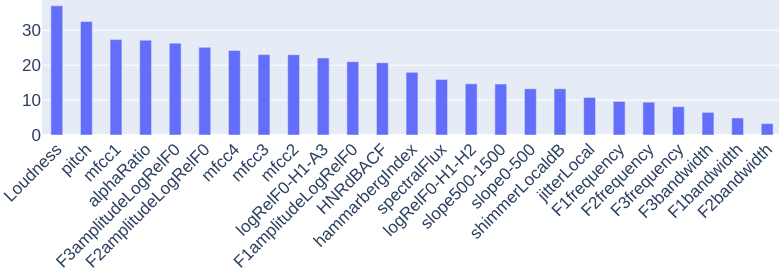} \\
    \caption{Acoustic improvement (in \%) for FullSubNet (upper) and Demucs (lower) by using the proposed PAAP Loss, where acoustic improvement is reduction in MAE as defined in Section \ref{subsec:acoustic-improvement}.}
    \label{fig:ac_improvement}
    \vspace{-5mm}
\end{figure}

\vspace{-3mm}
\subsection{Analysis}

\subsubsection{Acoustic improvement}
\label{subsec:acoustic-improvement}

Fig. \ref{fig:ac_improvement} provides a visualization of the percentage improvement of the 25 acoustic parameters after using the PAAP Loss to fine-tune the model. The acoustic improvement is measured by the reduction in mean absolute error (MAE) between the acoustic parameters of the enhanced and clean speech. Formally, if $\D^E, \D^C \in \mathbb{R}^{N \times 25} $ are the enhanced and clean estimated acoustics, for each acoustic parameter $j$ we compute 
\begin{equation}
\text{MAE}(\D^E_j, \D^C_j) = \frac{1}{N} \sum_{i=1}^N |\D^E_{ij} - \D^C_{ij} |
\end{equation}  
and then average over all acoustic parameters to get $\text{MAE}(\D^E, \D^C)$. 
Formally, the acoustic improvement as reduction in MAE is
\begin{equation}
\frac{\text{MAE}(\D^E, \D^C) - \text{MAE}(\D^B, \D^C)}{\text{MAE}(\D^B, \D^C)} \cdot 100\%
\end{equation}

where $\D^B$ stands for the acoustic parameters from the baseline enhancement model. For FullSubNet, we can observe that the PAAP Loss has the most improvement on MFCC features and loudness. On the other hand, for Demucs, most of the acoustic improvement of features are at the similar level with FullSubNet, except that the loudness and the F0 on a semitone frequency scale have a larger boost of ~30\%. Among all the acoustic features, the acoustic improvements are relatively small for formant frequencies and formant bandwidths for both models, but we conclude that we are getting a consistent improvement on all of the acoustic low-level descriptors across different categories of SE models.  

\subsubsection{Phoneme-dependent acoustic improvement}
\label{subsec:acoustic-phonetic-improvement}


In the previous section, we looked at overall improvements for each acoustic parameter. Now we break down the analysis further by showing the improvement in each acoustic parameter segmented by phoneme. The acoustic improvement is calculated by first creating phoneme alignments with the phonetic aligner on the clean speech. Then for each frame, we take the difference in acoustic parameters for clean and enhanced speech, and add this difference to the running total of the corresponding aligned phoneme. At the end, we average the differences per phoneme by the number of frames.   

\begin{figure}[t]
    \centering
    \includegraphics[width=0.8\columnwidth]{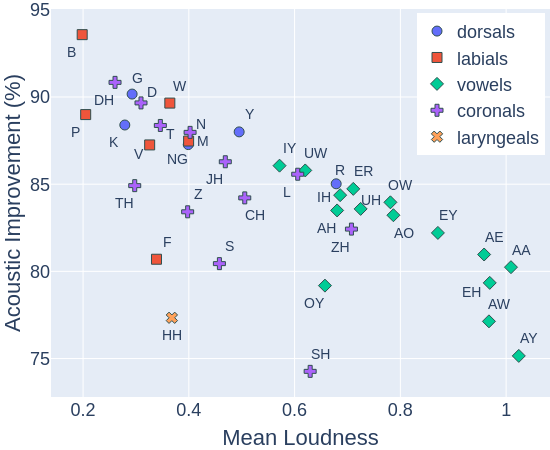} 
    
    \vspace{1.5mm}
    
    \includegraphics[width=0.8\columnwidth]{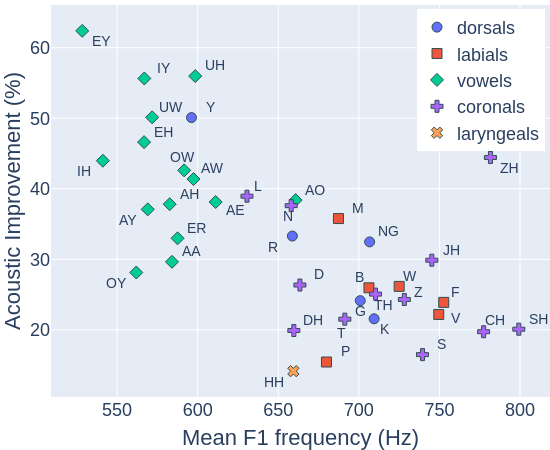}
    \caption{Reduction in error of loudness / F1 frequency vs. average value of acoustic parameter for each phoneme.}
    \label{fig:phone_improvement}
    \vspace{-5mm}
\end{figure}

We connect this analysis with the acoustic-phonetic properties mentioned in the introduction. Recall that plosives have very characteristic behavior with amplitude features. Also recall that vowels and nasals have specific formant characteristics. We include plots of per-phoneme acoustic parameter improvement for loudness and F1 frequency to represent the amplitude and formant characteristics, respectively. 

We plot the phoneme-dependent improvement for loudness and formant-1 (F1) frequency in Fig. \ref{fig:phone_improvement}. Each phoneme represents one point, where the colors/shapes indicate different phoneme categories. We separate out vowels, and then use the place of articulation as the classification standard of consonants. This includes dorsals, labials and coronals, which correspond to consonants where the articulation is performed with tongue dorsum, lips, and tongue front respectively. We also separate /HH/ as the only consonant in English with the place of articulation in the larynx. Therefore, we use five different colors/shapes in total to represent phoneme categories in the figure.
 
 With this knowledge, we can see that our phonetically-aligned acoustic parameter loss results in the expected improvements given the above domain knowledge. The highest improvements in loudness are in plosives such as /B/, /P/, /K/, /G/, /D/, and /DH/, where the average improvement is around 90\%. The goal of the PAAP Loss was to learn the relations between phonemes and acoustic parameters over time, and fine-tune enhancement models to account for this. Now we observe models fine-tuned with PAAP Loss produce speech with more improvement in acoustic parameters for the specific phonemes that are relevant for that particular parameter. 
 
 We also see the expected clustering of improvement for F1 frequency. Nearly all the highest improvements are seen with vowels, as formant structure is more important for vowels than consonants. The overall acoustic improvement of vowels is around 45\%, higher than any group of consonants. The nasals /N/ and /M/, also mentioned in the introduction for their formant structure, showed similar improvements to many vowels. The other consonants that showed high improvement, /L/ and /R/ are liquid consonants, which are known to be more similar to vowels than other consonants.

\vspace{-3mm}



\vspace{-1mm}
\section{Conclusion}
\vspace{-2.7mm}
\label{sec:conclusion}

In this work, we propose a novel auxiliary objective for speech enhancement, the phonetic-aligned acoustic parameter (PAAP) loss, which minimizes the differences between important temporal acoustic parameters that are weighted by phoneme types. We fine-tune competitive speech enhancement models with the addition of PAAP Loss, and experiments show that performance increases across all evaluation metrics, including measures of perceptual quality, and WER from competitive ASR models. We provide a detailed analysis of the phoneme-dependent acoustic improvement to show that the acoustic parameters improve most in expected phoneme categories.

\vspace{-1.5mm}
\section{Acknowledgement}
\vspace{-3mm}
This work used the Extreme Science and Engineering Discovery Environment (XSEDE) ~\cite{xsede}, which is supported by National Science Foundation grant number ACI-1548562. Specifically, it used the Bridges system ~\cite{nystrom2015bridges}, which is supported by NSF award number ACI-1445606, at the Pittsburgh Supercomputing Center (PSC).

\vfill\pagebreak


\section{References}
\printbibliography

@ARTICLE{regression_dnn_denoising,
  author={Xu, Yong and Du, Jun and Dai, Li-Rong and Lee, Chin-Hui},
  journal={IEEE/ACM Transactions on Audio, Speech, and Language Processing}, 
  title={A Regression Approach to Speech Enhancement Based on Deep Neural Networks}, 
  year={2015},
  volume={23},
  number={1},
  pages={7-19},
  doi={10.1109/TASLP.2014.2364452}}

@INPROCEEDINGS{rnn_denoising,
  author={Weninger, Felix and Eyben, Florian and Schuller, Björn},
  booktitle={Proc. ICASSP}, 
  title={Single-channel speech separation with memory-enhanced recurrent neural networks}, 
  year={2014},
  volume={},
  number={},
  pages={3709-3713},
  doi={10.1109/ICASSP.2014.6854294}}

@inproceedings{Weninger2015SpeechEW,
  title={Speech Enhancement with {LSTM} Recurrent Neural Networks and its Application to Noise-Robust {ASR}},
  author={Felix Weninger and Hakan Erdogan and Shinji Watanabe and Emmanuel Vincent and Jonathan Le Roux and John R. Hershey and Bj{\"o}rn Schuller},
  booktitle={LVA/ICA},
  year={2015}
}

@inproceedings{cnn_denoising,
  title={Convolutional-recurrent neural networks for speech enhancement},
  author={Zhao, Han and Zarar, Shuayb and Tashev, Ivan and Lee, Chin-Hui},
  booktitle={Proc. ICASSP},
  pages={2401--2405},
  year={2018},
  organization={IEEE}
}

@article{dnsmos_pesq_flaw,
  title={A Scalable Noisy Speech Dataset and Online Subjective Test Framework},
  author={Reddy, Chandan KA and Beyrami, Ebrahim and Pool, Jamie and Cutler, Ross and Srinivasan, Sriram and Gehrke, Johannes},
  journal={Proc. Interspeech},
  pages={1816--1820},
  year={2019}
}

@inproceedings{metricgan,
  title={Metricgan: Generative adversarial networks based black-box metric scores optimization for speech enhancement},
  author={Fu, Szu-Wei and Liao, Chien-Feng and Tsao, Yu and Lin, Shou-De},
  booktitle={International Conference on Machine Learning},
  pages={2031--2041},
  year={2019},
  organization={PMLR}
}

@article{yi2019encoding,
  title={The encoding of speech sounds in the superior temporal gyrus},
  author={Yi, Han Gyol and Leonard, Matthew K and Chang, Edward F},
  journal={Neuron},
  volume={102},
  number={6},
  pages={1096--1110},
  year={2019},
  publisher={Elsevier}
}

@article{alwan2011perception,
  title={Perception of place of articulation for plosives and fricatives in noise},
  author={Alwan, Abeer and Jiang, Jintao and Chen, Willa},
  journal={Speech communication},
  volume={53},
  number={2},
  pages={195--209},
  year={2011},
  publisher={Elsevier}
}

@inproceedings{plantinga2020phonetic,
  title={Phonetic feedback for speech enhancement with and without parallel speech data},
  author={Plantinga, Peter and Bagchi, Deblin and Fosler-Lussier, Eric},
  booktitle={Proc. ICASSP},
  pages={6679--6683},
  year={2020},
  organization={IEEE}
}

@article{plantinga2021perceptual,
  title={Perceptual Loss with Recognition Model for Single-Channel Enhancement and Robust {ASR}},
  author={Plantinga, Peter and Bagchi, Deblin and Fosler-Lussier, Eric},
  journal={arXiv preprint arXiv:2112.06068},
  year={2021}
}

@article{correlates_breathy,
author = {Hillenbrand, James and Cleveland, Ronald and Erickson, Robert},
year = {1994},
month = {09},
pages = {769-78},
title = {Acoustic Correlates of Breathy Vocal Quality},
volume = {37},
journal = {Journal of speech and hearing research},
doi = {10.1044/jshr.3704.769}
}

@article{correlates_breathy_rough,
  title={Some spectral correlates of pathological breathy and rough voice quality for different types of vowel fragments},
  author={Krom, Guus de},
  journal={Journal of Speech, Language, and Hearing Research},
  volume={38},
  number={4},
  pages={794--811},
  year={1995},
  publisher={ASHA}
}

@article{defossez2020real,
  title={Real time speech enhancement in the waveform domain},
  author={Defossez, Alexandre and Synnaeve, Gabriel and Adi, Yossi},
  journal={arXiv preprint arXiv:2006.12847},
  year={2020}
}

@inproceedings{hao2021fullsubnet,
  title={FullSubNet: a full-band and sub-band fusion model for real-time single-channel speech enhancement},
  author={Hao, Xiang and Su, Xiangdong and Horaud, Radu and Li, Xiaofei},
  booktitle={Proc. ICASSP},
  pages={6633--6637},
  year={2021},
  organization={IEEE}
}

@inproceedings{zhu2022phone,
  title={Phone-to-audio alignment without text: A Semi-supervised Approach},
  author={Zhu, Jian and Zhang, Cong and Jurgens, David},
  booktitle={Proc. ICASSP},
  pages={8167--8171},
  year={2022},
  organization={IEEE}
}

@article{reddy2020interspeech,
  title={The INTERSPEECH 2020 Deep Noise Suppression Challenge: Datasets, Subjective Testing Framework, and Challenge Results},
  author={Reddy, Chandan KA and Gopal, Vishak and Cutler, Ross and Beyrami, Ebrahim and Cheng, Roger and Dubey, Harishchandra and Matusevych, Sergiy and Aichner, Robert and Aazami, Ashkan and Braun, Sebastian and others},
  journal={Proc. Interspeech},
  year={2020}
}

@inproceedings{dubey2022icassp,
  title={{ICASSP} 2022 deep noise suppression challenge},
  author={Dubey, Harishchandra and Gopal, Vishak and Cutler, Ross and Aazami, Ashkan and Matusevych, Sergiy and Braun, Sebastian and Eskimez, Sefik Emre and Thakker, Manthan and Yoshioka, Takuya and Gamper, Hannes and others},
  booktitle={Proc. ICASSP},
  pages={9271--9275},
  year={2022},
  organization={IEEE}
}

@inproceedings{graz_university,
  title={A pitch tracking corpus with evaluation on multipitch tracking scenario},
  author={Pirker, Gregor and Wohlmayr, Michael and Petrik, Stefan and Pernkopf, Franz},
  booktitle={Proc. Interspeech},
  year={2011}
}

@inproceedings{reddy2022dnsmos,
  title={{DNSMOS} {P}. 835: A non-intrusive perceptual objective speech quality metric to evaluate noise suppressors},
  author={Reddy, Chandan KA and Gopal, Vishak and Cutler, Ross},
  booktitle={Proc. ICASSP},
  year={2022},
  organization={IEEE}
}

@inproceedings{manocha2022noresqa,
  title={Speech Quality Assessment through MOS using Non-Matching References},
  author={Manocha, Pranay and Kumar, Anurag},
  booktitle={Proc. Interspeech},
  year={2022}
}

@inproceedings{yang2022improving,
  title={Improving Speech Enhancement through Fine-Grained Speech Characteristics},
  author={Yang, Muqiao and Konan, Joseph and Bick, David and Kumar, Anurag and Watanabe, Shinji and Raj, Bhiksha},
  booktitle={Proc. Interspeech},
  year={2022}
}

@inproceedings{taploss,
  title={{TAPLoss}: A Temporal Acoustic Parameter Loss for Speech Enhancement},
  author={Zeng, Yunyang and Konan, Joseph and Han, Shuo and Bick, David and Yang, Muqiao and Kumar, Anurag and Watanabe, Shinji and Raj, Bhiksha},
  booktitle={Proc. ICASSP},
  year={2023}
}

@article{chen2022wavlm,
  title={{WavLM}: Large-scale self-supervised pre-training for full stack speech processing},
  author={Chen, Sanyuan and Wang, Chengyi and Chen, Zhengyang and Wu, Yu and Liu, Shujie and Chen, Zhuo and Li, Jinyu and Kanda, Naoyuki and Yoshioka, Takuya and Xiao, Xiong and others},
  journal={IEEE Journal of Selected Topics in Signal Processing},
  year={2022},
  publisher={IEEE}
}

@article{egemaps,
  title={The Geneva minimalistic acoustic parameter set ({GeMAPS}) for voice research and affective computing},
  author={Eyben, Florian and Scherer, Klaus R and Schuller, Bj{\"o}rn W and Sundberg, Johan and Andr{\'e}, Elisabeth and Busso, Carlos and Devillers, Laurence Y and Epps, Julien and Laukka, Petri and Narayanan, Shrikanth S and others},
  journal={IEEE transactions on affective computing},
  volume={7},
  number={2},
  pages={190--202},
  year={2015},
  publisher={IEEE}
}

@ARTICLE{xsede,
author = {J. Towns and T. Cockerill and M. Dahan and I. Foster and K. Gaither and A. Grimshaw and V. Hazlewood and S. Lathrop and D. Lifka and G. D. Peterson and R. Roskies and J. R. Scott and N. Wilkins-Diehr},
journal = {Computing in Science \& Engineering},
title = {XSEDE: Accelerating Scientific Discovery},
year = {2014},
volume = {16},
number = {5},
pages = {62-74},
doi = {10.1109/MCSE.2014.80},
url = {doi.ieeecomputersociety.org/10.1109/MCSE.2014.80},
ISSN = {1521-9615},
month={Sept.-Oct.}
}

@inproceedings{nystrom2015bridges,
  title={Bridges: a uniquely flexible HPC resource for new communities and data analytics},
  author={Nystrom, Nicholas A and Levine, Michael J and Roskies, Ralph Z and Scott, J Ray},
  booktitle={Proceedings of XSEDE Conference: Scientific Advancements Enabled by Enhanced Cyberinfrastructure},
  pages={1--8},
  year={2015}
}

@article{KASUYA1986171,
title = {An acoustic analysis of pathological voice and its application to the evaluation of laryngeal pathology},
journal = {Speech Communication},
doi = {https://doi.org/10.1016/0167-6393(86)90006-3},
url = {https://www.sciencedirect.com/science/article/pii/0167639386900063},
author = {Hideki Kasuya and Shigeki Ogawa and Yoshinobu Kikuchi and Satoshi Ebihara},
  year={1986}
}

@article{nasals,
author = {Styler,Will },
title = {On the acoustical features of vowel nasality in English and French},
journal = {The Journal of the Acoustical Society of America},
volume = {142},
number = {4},
pages = {2469-2482},
year = {2017},
doi = {10.1121/1.5008854},

URL = { 
        https://doi.org/10.1121/1.5008854
    
},
eprint = { 
        https://doi.org/10.1121/1.5008854
   
}
}

@article{turian_henry,
  title={I'm sorry for your loss: Spectrally-based audio distances are bad at pitch},
  author={Turian, Joseph and Henry, Max},
  journal={arXiv preprint arXiv:2012.04572},
  year={2020}
}

@article{perceptual_jnd,
  title={A differentiable perceptual audio metric learned from just noticeable differences},
  author={Manocha, Pranay and Finkelstein, Adam and Zhang, Richard and Bryan, Nicholas J and Mysore, Gautham J and Jin, Zeyu},
  journal={Proc. Interspeech},
  year={2020}
}

@article{metric_gan_plus,
  title={Metricgan+: An improved version of metricgan for speech enhancement},
  author={Fu, Szu-Wei and Yu, Cheng and Hsieh, Tsun-An and Plantinga, Peter and Ravanelli, Mirco and Lu, Xugang and Tsao, Yu},
  journal={Proc. Interspeech},
  year={2021}
}

@ARTICLE{LSTM,
  author={Hochreiter, Sepp and Schmidhuber, Jürgen},
  journal={Neural Computation}, 
  title={Long Short-Term Memory}, 
  year={1997},
  volume={9},
  number={8},
  pages={1735-1780},
  doi={10.1162/neco.1997.9.8.1735}}

@INPROCEEDINGS{metric_RL_SE,
  author={Koizumi, Yuma and Niwa, Kenta and Hioka, Yusuke and Kobayashi, Kazunori and Haneda, Yoichi},
  booktitle={Proc. ICASSP}, 
  title={{DNN}-based source enhancement self-optimized by reinforcement learning using sound quality measurements}, 
  year={2017},
  volume={},
  number={},
  pages={81-85},
  doi={10.1109/ICASSP.2017.7952122}}

@ARTICLE{white_box_perceptual_loss,
  author={Martin-Doñas, Juan Manuel and Gomez, Angel Manuel and Gonzalez, Jose A. and Peinado, Antonio M.},
  journal={IEEE Signal Processing Letters}, 
  title={A Deep Learning Loss Function Based on the Perceptual Evaluation of the Speech Quality}, 
  year={2018},
  volume={25},
  number={11},
  pages={1680-1684},
  doi={10.1109/LSP.2018.2871419}}

@article{hsieh2020improving,
  title={Improving perceptual quality by phone-fortified perceptual loss using {Wasserstein} distance for speech enhancement},
  author={Hsieh, Tsun-An and Yu, Cheng and Fu, Szu-Wei and Lu, Xugang and Tsao, Yu},
  journal={Proc. Interspeech},
  year={2021}
}

@inproceedings{tal22_interspeech,
  author={Or Tal and Moshe Mandel and Felix Kreuk and Yossi Adi},
  title={{A Systematic Comparison of Phonetic Aware Techniques for Speech Enhancement}},
  year=2022,
  booktitle={Proc. Interspeech},
  pages={1193--1197},
  doi={10.21437/Interspeech.2022-695}
}

@article{hsu2021hubert,
  title={Hubert: Self-supervised speech representation learning by masked prediction of hidden units},
  author={Hsu, Wei-Ning and Bolte, Benjamin and Tsai, Yao-Hung Hubert and Lakhotia, Kushal and Salakhutdinov, Ruslan and Mohamed, Abdelrahman},
  journal={IEEE/ACM Transactions on Audio, Speech, and Language Processing},
  volume={29},
  pages={3451--3460},
  year={2021},
  publisher={IEEE}
}

@inproceedings{gemmeke2017audio,
  title={Audio set: An ontology and human-labeled dataset for audio events},
  author={Gemmeke, Jort F and Ellis, Daniel PW and Freedman, Dylan and Jansen, Aren and Lawrence, Wade and Moore, R Channing and Plakal, Manoj and Ritter, Marvin},
  booktitle={Proc. ICASSP},
  pages={776--780},
  year={2017},
  organization={IEEE}
}

\end{document}